\documentclass{ws-ijmpa}
\usepackage[super,compress]{cite}
\usepackage[utf8]{inputenc}
\usepackage{amsmath,amsxtra,amssymb,latexsym,amscd,mathtools,tensor}
\usepackage{graphicx,wrapfig,subfigure,setspace}
\usepackage[usenames,dvipsnames]{color}
\usepackage{array,comment}
\usepackage{url}
\usepackage[citecolor=blue, colorlinks = true]{hyperref}
\usepackage{tabu}

\begin{document}

\markboth{V. Lukovi\'{c}, P. Cabella, N. Vittorio}{Dark Matter Review}

\catchline{}{}{}{}{}

\title{DARK MATTER IN COSMOLOGY}

\author{VLADIMIR LUKOVI\'{C}}
\address{University Tor Vergata, Dipartimento di Fisica, Gruppo di cosmologia,\\
Via della ricerca scientifica 1, Roma, 00133, Italy
%\\first\_author@domain\_name
}

\author{PAOLO CABELLA}
\address{University Tor Vergata, Dipartimento di Fisica, Gruppo di cosmologia,\\
Via della ricerca scientifica 1, Roma, 00133, Italy\\
paolo.cabella@roma2.infn.it}

\author{NICOLA VITTORIO}
\address{University Tor Vergata, Dipartimento di Fisica, Gruppo di cosmologia,\\
Via della ricerca scientifica 1, Roma, 00133, Italy\\
nicola.vittorio@roma2.infn.it}

\maketitle

\begin{abstract}
In this paper we review the main theoretical and experimental achievements in the field of Dark Matter from the Cosmological and Astrophysical point of view. We revisit it from the very first surveys of local astrophysical matter, up to the stringent constraints on matter properties, coming from the last release of data on cosmological scales. 
To bring closer and justify the idea of dark matter, we will go across methods and tools for measuring dark matter characteristics, and in some cases a combination of methods that provide one of the greatest direct proofs for dark matter, such as Bullet cluster.
\end{abstract}
\section{Introduction}
Historically, the first hints of dark matter from astrophysical point of view date back in first half of the last century. They came from the first trials to estimate matter density in our galactic neighbourhood.\\
Estonian astronomer Ernst \"{O}pik \cite{Opik15} and Dutch astronomer Jacobus Kapteyn \cite{Kapteyn22} made first attempts on estimating the density of matter in vicinity of the Solar system by analyzing arrangement and vertical motions of nearby stars. By comparing this dynamical estimate of matter density to total density due to all stars near the Galactic
plane using the luminosity function of stars, \"{O}pik and Kapteyn found that the surface density of known stars in Galactic plane is sufficient to explain their vertical motions. On the other hand, British astronomer James Jeans \cite{Jeans22} introduced some corrections to Kapteyn's model, reanalyzed vertical motions of stars and made better estimate of matter density, with conclusion that some extra matter probably exists near our system. In fact, Jeans indicates that on average two dark stars are required per every bright star.
\\
Further development of the topic was done by another Dutch astronomer Jan Oort \cite{Oort32}. He also made good estimate of density contribution due to faint stars (i.e. not observed) that was enough to explain the lack in dynamical matter density.
\\
Term Dark Matter, `Dunkle Materie', (DM), in the sense in which it is used today, was firstly introduced by Swiss astronomer Fritz Zwicky in early 1930s. He was the first one to apply virial theorem to estimate the total mass of large structure, particularly of Coma cluster \cite{Zwicky33}. Zwicky measured the variance of peculiar velocities of visible matter in Coma cluster, via redshift, and showed that relative speeds of galaxies in Coma were much too great for them to be held together by the gravitational attraction of the visible matter alone, and that therefore, there must have been something else holding them together. His observations suggest the total gravitational mass in Coma cluster is about two orders of magnitude larger than the visible mass. He also suggested other new methods for measuring total mass of large structures like gravitational lensing \cite{Zwicky37}.
\\
In the following years more stringent evidences of dark matter were coming. American astronomer Horace Babcock and Oort obtained and analyzed rotational curves of stars in a galaxy. This method is currently used for measuring DM distribution. In his PhD thesis \cite{Babcock39}, Babcock reported measurements of the rotation curve for Andromeda (M31) that are flat on the periphery, instead of expected Keplerian decrease due to rapid fall of luminosity on the periphery. Oort encountered on the similar problems when analyzing two elliptical galaxies \cite{Oort40}. This results suggested the mass-to-luminosity ratio increases radially inside galaxy. Ignoring, however, already many evidences for enormous mass of non-visible matter, Babcock attributed this `measurement problem' in rotational curves to other effects. However, similar behavior was soon observed in other spiral galaxies (see section \ref{rotcur}).
\\
More methods are being proposed for measuring mass of large scale structures. American astronomer Thornton Page estimates mass of systems of double elliptical galaxies by analyzing their dynamics, showing the lack of mass compared to luminosity \cite{Page52, Page59, Page60}. Analogously, British astronomer Franz Kahn and an Oort's student L. Woltjer made another estimate of mass of M31 and Milky Way. They measured the relative attraction velocity from the blueshift of M31. Considering two galaxies as a closed system, knowing their present distance and taking the age of the Universe as an upper limit for the time of attraction they deduced that the effective mass was at least several times larger than the sum of M31 and Milky Way visible mass \cite{KahnWoltjer59}.\\
Still the idea about dark matter was not universally accepted in astronomical community. At that time somewhat controversial findings got Russian astronomer Grigori Kuzmin, who was doing a research on star dynamics in 1950s with his students. Using Oort's method and more data on stars' vertical motions, they report that no dark matter in Galactic disk near Solar system is needed \cite{Kuzmin52, Kuzmin55}.
\section{Direct velocity measurements}
Velocity fields and methods for analyzing dark matter in galaxies strongly depend on galaxy type and its physical properties. Stellar population in spiral galaxies rotates in ordered manner with a differential velocity field as a well-known disk structure. On the other hand, elliptical galaxies have three dimensional random velocity field well approximated by Maxwell distribution. The main observable for the formers is the reconstruction of the rotational curves, while for the later is the measurement dispersion velocity distribution via Virial theorem.
\subsection{Rotational curves}
\label{rotcur}
Vesto Slipher working at Lowell Observatory, USA, made the first demonstration that spiral galaxies (nebulae) rotate by detecting inclined absorption lines in the nuclear spectra of Andromeda and Sombrero galaxies \cite{Slipher14}. In years that followed the same has been shown for other nebulae, but Babcock was the first to extend the measurements of rotational curves further from the nucleus. These early results were very inaccurate, though it was enough to note that the outer parts of the disk were rotating with unexpectedly high velocities, consequently raising questions of mass distribution. Concretely, assuming even circular ordered rotation of the spiral galaxy with differential velocity $v(r)$ and using Newton's law of gravitation, it is simply to find relation of velocity at radius $r$ and density or mass contained in a shell $M(r)$: ${\rm G}M(r)=r v^2$. One would expect that after the luminous disk ends, velocity curve $v(r)$ falls, but that is not what was observed. As Oort describes: ``...the distribution of mass appears to bear almost no relation to that of the light." When analyzing rotation curve of NGC 3115, he finds that in outer regions ratio of mass to light, in solar units, is about 250, which is two orders of magnitude larger than in the neighborhood of the solar system \cite{Oort40}.\\
Others continued measuring and extending rotational curves, also developing new methods. In time, instruments developed enabling astronomers to distinguish separate components in the spiral galaxy structure, like nucleus, core, bulge, disk, spiral arms and halo. It is important to say that different methods had to be used for extending rotational curves through different regions. A summary on early development of methods can be obtained from work of Jaan Einasto et. al. \cite{Einasto74} and references there in. This Estonian astronomer is also famous namely for analyzing motions of satellite galaxies as a method for measuring properties of the halo component, including measurement of rotational curves in this region. Motions of satellite galaxies showed that mass and radius of halo is an order of magnitude larger than visible disk \cite{Ostriker74}. In early 70s, the mean density of Universe was estimated to be about 0.2 of critical density, that is 4-5 times more than believed earlier, due to amount of DM.\\
Many have demonstrated that spiral galaxies do not spin as expected according to Keplerian dynamics, but the major breakthrough came from American astronomers Vera Rubin and K. Ford, when they started using a new sensitive spectrograph. Rubin worked on rotational curves from the late 60s, through 70s and later. Most of her results are collected in two influential papers \cite{Rubin80, Rubin85}. Overall conclusion was that ``rotation curves of high luminosity spiral galaxies are flat, at nuclear distances as great as 50 kpc" \cite{Rubin78}. Radio observations of the 21 cm line of HI, increased resolution and also extended image of the galaxies, revealing that rotation curves remain flat ($v(r)\sim const$), even beyond the optical disks \cite{Bosma78}.\\
Rotational curves suggest that more than half of the mass of galaxies is contained in the relatively dark galactic halo, or, that Newtonian gravitational theory does not hold over astrophysical distances. For this reason alternative theories of gravity were proposed to try to explain the anomalies for which dark matter is intended to account. One of the most discussed models is Modified Newtonian Dynamics (MOND), proposed soon after the experimental findings \cite{Milgrom83}, and developed ever since. Similar alternative has been proposed recently \cite{Moffat06}.\\
The conservative view is today even more accepted by scientific community, because of discovering more, and new types of evidence. It leads to the difference between the galaxy mass predicted by the luminosity and the mass predicted by the velocities. This difference offers strong evidence that spiral galaxies are embedded in extended halos of dark matter. Comparison of density profile to luminosity profile is better seen from a radial dependance of the mass-to-luminosity ratio ($M / L$), and it is a clue to the distribution of both visible and invisible mass.\\
Assuming either spherical or flat disk distribution of mass, one can derive mass density distribution from rotational curve as shown by Takamiya and Sofue \cite{Takamiya00}, who used formulae presented by Binney and Tremaine \cite{Binney87}. Including luminosity profiles one can also calculate distribution of the $M/L$. Increment of $M/L$ indicates presence of dark matter in the region. However, for fitting the experimental data, a model profile is needed. Several have been suggested: Einasto \cite{Einasto65}, Burkert \cite{Burkert95}, or the pseudo-isothermal profile, but the best known is Navarro-Frenk-White (NFW) profile. The latter one is fitted to dark matter haloes identified in N-body simulations of large scale structures \cite{NFW96}, and it reads:
\begin{equation}
\label{NFWpf}
\rho (r)=\frac{\rho_0}{\frac{r}{R_s}\left(1~+~\frac{r}{R_s}\right)^2},
\end{equation}
where $\rho_0$ and ``scale radius", $R_s$, are free parameters. NFW is the most commonly used profile for dark matter halos. 
\begin{figure}
\center
\includegraphics[scale=0.7]{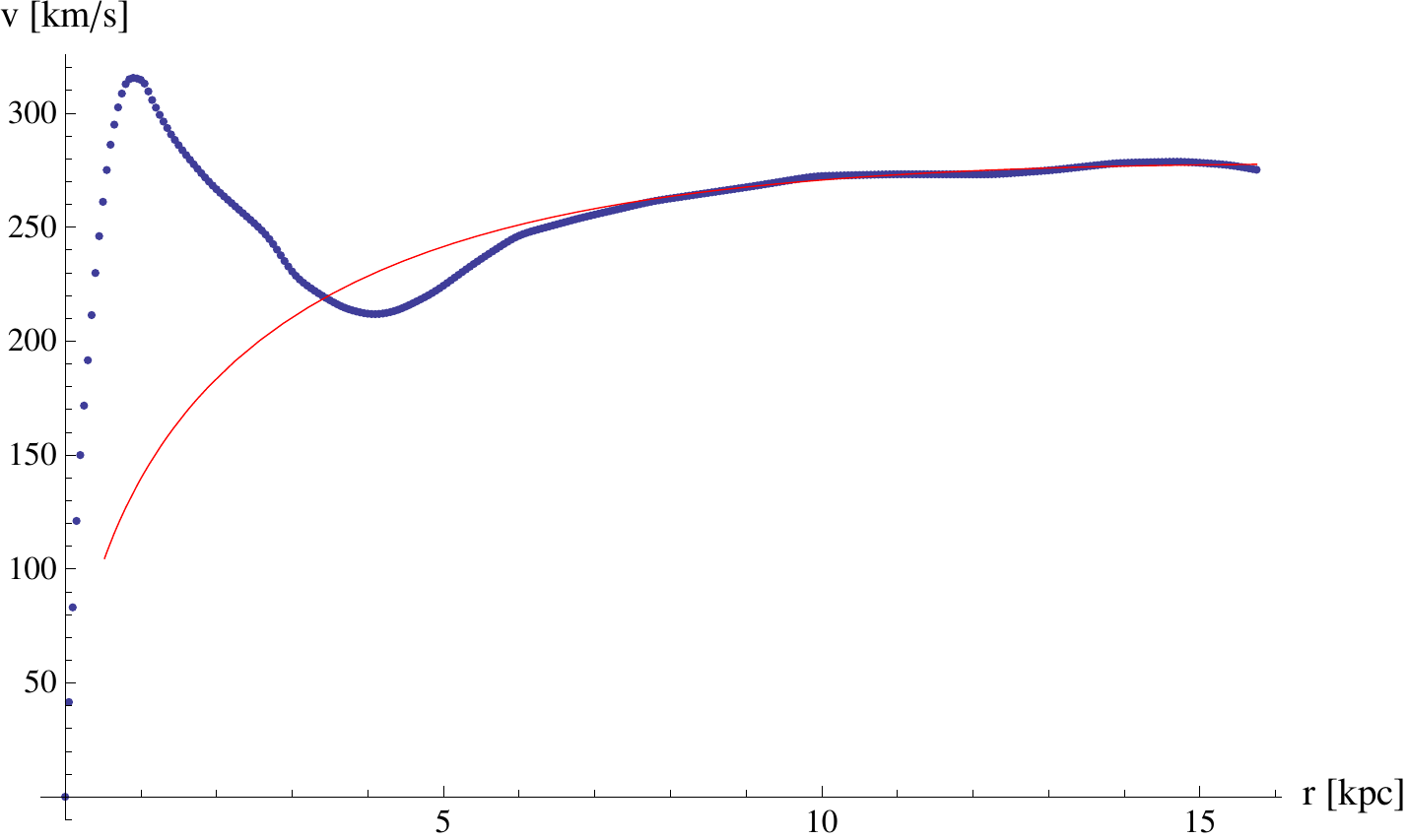}
\caption{Rotational curve of NGC2590 (blue), fitted to NFW profile for the halo (red).}
\label{NFWplot}
\end{figure}
On figure \ref{NFWplot}. we show rotational curve of NGC 2590, fitted with NFW profile in the part of the halo. Simple NFW profile, Eq. \eqref{NFWpf} has limited application to the region of halo component, while more general profiles have five or more parameters.
\\
Another simple procedure to obtain density profile is to assume that $ M / L$ is  constant throughout each galactic component, like bulge and disk \cite{Kent86}.\\
There were even some discussions that dark matter halo is needed for stability of disk structure in Spirals \cite{Ostriker73} (but check also \cite{Sellwood85}). The problem of distribution of dark matter was far from being resolved. One of uncertainties still is the question of existence of distinctive separation between  the disk (luminous or dark) and the halo. The size and the shape of the halo in double systems is another tricky question. For example, a Milky Way halo extends at least 200 kpc and it is getting close to the half-way distance between the Galaxy and Andromeda, 350 kpc. And if halos are as large as those suggested by the gravitational distortion of background galaxies seen in the vicinity of foreground, then the halo of our Galaxy may brush the equivalently large halo of M31. In addition, few spiral galaxies exhibit a true Keplerian decline in their rotation velocities.\\ %{\bf(put some new references)}
Few elliptical galaxies are surrounded by the ring of H{\rm I} gas in the outermost regions. Rotation of this gas can be traced and used just like rotation curves in spiral galaxies, as a method to study the dark matter halo \cite{Schweizer89, Bertola93}.
\subsection{Virial evidences}
Velocity dispersion of stellar population in elliptical galaxy, or equivalently, of galaxies in a cluster determines the kinetic energy of the structure. Assuming that the structure is in virial equilibrium, we are able to estimate its total mass, $M$, from the virial theorem. In a simple approach, for spherical structure, virial theorem states:
\begin{equation}
\frac{{\rm G} M}{R}=\sigma^2,
\end{equation}
where $R$ is the radius of structure, inside of which velocity dispersion $\sigma$ is measured. Comparing this to total luminosity allows us to make conclusions on the amount of dark matter in the structure. Applying the method to the central region of elliptical galaxy requires no dark matter to explain central brightness \cite{Binney90}. However, Bertin and Saglia in 1990 developed spherical, collisionless, self-consistent two-component model (with a halo), that was then used on a number of elliptical galaxies and was able to recognize the presence of dark matter of the order of luminous mass, in the form of a dark halo \cite{Bertin92}.\\
One can also analyze radial distribution of velocity dispersion in the structure, and relate it to matter distribution \cite{Saglia93}. Unlike rotation curves in spirals, there is big variety of different dispersion profiles. Overall conclusion for elliptical galaxies is that amount of dark matter in the luminous region is certainly not the dominating mass component \cite{Paolis95}. Some Ellipticals show flat or quasi-flat dispersion curve, indicating presence of dark halos \cite{Carollo95, Bertin94}, but it is not always the case, which makes our understanding of the structure of Ellipticals even more problematic (see for example \cite{Morganti13}).\\
 In this approach one can use Jeans equations:
\begin{equation}
\frac1\rho\frac{d(\rho\sigma_r^2)}{dr}+2\frac{\beta\sigma_r^2}{r}+\frac{d\phi}{dr}=0
\label{Jeans}
\end{equation}
where $\rho(r)$ is only stellar mass density, $\phi(r)$ is total gravitational potential, $\beta(r)$ is the velocity dispersion anisotropy,
\begin{equation}
\beta(r)=1-\frac{\left<v_\theta^2\right>}{\left<v_r^2\right>},
\end{equation}
$v_\theta$ and $v_r$ being azimuthal and radial velocity components. Anisotropy parameter $\beta(r)$ measures the overall shape of orbits of stars(galaxies) in the structure. Value $\beta=-\infty$, if the orbits are perfectly circular; and $\beta=1$, if orbits are fully radial. Unfortunately, one can not measure all the unknowns from Eq. \eqref{Jeans}, and velocities are measured only along the line of sight. Solution is to assume model functions for the two dispersion components in $\beta(r)$ with a number of free parameters that are fitted to measured data points, see \cite{Napolitano11}. Napolitano et. al. did confirm the existence of huge dark matter halos in Ellipticals. Distribution $\rho(r)$ can be obtained from measurements, but it is commonly enhanced to correct for very faint objects.
\section{X-ray observations}
The most promising method to study dark matter in Ellipticals and clusters of galaxies is, in fact, the analysis of the interstellar, or intracluster medium (ICM), respectfully. They consist of hot gas typically  with temperatures of ${\rm T}\sim10^7-10^8$K, dominantly radiating in X-ray, which is even the dominating baryonic component in both structures, an order of magnitude larger than stellar population.\\ %(maybe also larger than mass of all galaxies including their dark matter halos - check)
The X-ray emission of this hot gas is mainly due to thermal bremsstrahlung and line emission \cite{Binney87}. There are smaller contributions to continuum from recombination process and from two-photon decays of 2s levels in hydrogenic and helium-like ions \cite{Sarazin08}.\\
By studying the distribution and temperature of this hot gas it is possible to measure the gravitational potential of a galaxy or a cluster. This allows the determination of the total mass contained in the quoted objects. The exact study procedure is, theoretically, a simple one. For bremsstrahlung process, the emissivity of the gas, and hence, the brightness is proportional to the electron and proton number densities, or to the square of mass density of the gas. Measured X-ray brightness profile reflects the gas distribution profile, which is bound to gravitational potential profile of a structure. Results clearly show that visible mass alone is insufficient to cover for the total gravitational potential.\\
It is reasonable to consider this gas to be in hydrostatic equilibrium, since relaxation time is of the order of sound wave period through the structure, which is at these temperatures comparable to the orbital periods. Hydrostatic equilibrium relates pressure of the gas $p(r)$ with the total gravitational potential or total mass contained in a shall $M(r)$:
\begin{equation}
\label{HE}
{\mbox{d} p(r) \over \mbox{d} r} = - {\rm G}{M(r) \rho(r) \over r^2 }
\end{equation}
Substituting pressure from the equation for ideal gas $p=n{\rm k}T=\frac{\rho}{\mu m_p}{\rm k}T$, it becomes:
\begin{equation}
\label{Mtot}
M(r)=-\frac{k T(r)r}{\mu m_p G}\left[{\mbox{d} \ln\rho(r) \over \mbox{d}\ln r} + {\mbox{d} \ln T(r) \over \mbox{d}\ln r}\right],
\end{equation}
Another common assumption is the simple isothermal model, which might result from effective heat conduction via free electrons. Generally, temperature is calculated from the line ratios or from the continuum part in the X-ray spectra. Dark matter distribution is usually modeled with King model \cite{King72}:
\begin{equation}
\label{King}
\rho_{\rm DM}(r)=\rho_{\rm DM,0}(r)\left[1+\left(\frac r {r_c}\right)^2\right]^{-3/2},
\end{equation}
where $r_c$ is the core radius and $\rho_{\rm DM,0}$ is the central DM density. Gas density distribution is directly related to the X-ray brightness profile. Considering DM as main mass component, and assuming isothermal conditions, brightness and density profiles of the gas can be written in $\beta$-model
\begin{align}
\label{beta}
I(r)&=I_0(r)\left[1+\left(\frac r {r_c}\right)^2\right]^{-3\beta+1/2}\\
\rho(r)&=\rho_0(r)\left[1+\left(\frac r {r_c}\right)^2\right]^{-3\beta/2}
\end{align}
Parameters of the model are fitted from observations while temperature is obtained from the spectra, so that Eq. \eqref{Mtot} can finally be used to determine total mass.
\\
From the earliest X-ray observations it was clear that hot gas in clusters does not have sufficient mass \cite{Forman72, Gursky72, Kellogg73}.
In case of Ellipticals, an interesting example is M87. Fabricant et al. (1980) \cite{Fabricant80} found a value for the total mass $M\sim1.7-2.4\times 10^{13}{\rm M}_\odot$, while the gas mass was about $M_{gas}\sim10^{12}{\rm M}_\odot$. Similar behavior is observed in small galaxies groups \cite{David95}. More recent values show that $M_{gas}\approx6M_{gal}$ at large radii \cite{Voevodkin04}. It appears that most of the baryons today are in Warm Hot Intergalactic Medium (WHRM). However, DM highly surpasses even this: clusters of galaxies are DM $\sim84\%$, hot gas $\sim14\%$ and only few percents stars and galaxies \cite{Allen04}.
%Dark matter in Solar system \cite{Hill60}, \cite{Oort60}, \cite{Bahcall84}, \cite{Gilmore89}.
\section{Gravitational lensing}
\begin{figure}[h]
\center
\includegraphics[scale=0.7]{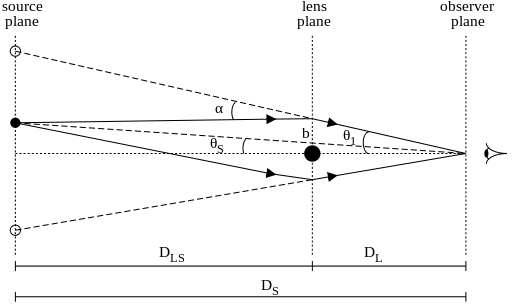}
\caption{Scheme of gravitational lensing}
\label{lens}
\end{figure}

Another evidence for DM existence and another method of measuring matter distribution in galaxies and clusters comes from gravitational lensing. For studying gravitational lensing in the beginning the most luminous clusters were selected. However, in modern times there are many clusters detected solely by lensing effects \cite{Wittman06}. Recent observations of Bullet cluster reveal separation of visible matter, ICM and dark matter halo, which were individually located using X-ray observations and lensing technique. This is why Bullet cluster is often cited as one of the best astrophysical evidences for dark matter models.\\
Theoretically, effect of gravitational deflection of light, was firstly pointed out at the end of 18th century, independently by British, Henry Cavendish, and German, Johann G. von Soldner, physicists \cite{Soldner804}. They suggested that the light ray should bend when passing close to massive celestial objects. With introduction of General relativity, Einstein recalculated the deflection angle in simple example of point-like objects. The correct value was shown to be two times larger than the one previously calculated in Newtonian mechanics \cite{Einstein15}:
\begin{equation}
\alpha = \frac{4GM}{b\,c^2}.
\end{equation}
Here, $\alpha$ is the angle of deflection, $M$ is the mass of gravitational lens, and $b$ is the impact parameter of the bended light ray (figure \ref{lens}).
\subsection{Strong lensing}
Gravitational deflection of light makes distorted image of objects located behind the lens. In case of strong gravitational lensing, multiple images of lensed object or arcs are created. First detection of gravitational lensing was in 1979; the Twin quasar was observed as two images of the same quasar created by strong lensing effect from massive elliptical galaxy \cite{Walsh79}. It is probably the best studied gravitationally lensed object. A particular example of strong lensing is the appearance of Einstein's ring. The effect was first discussed by Chwolson in 1924. \cite{Chwolson24}, and later remarked by Einstein \cite{Einstein36}. If the source, lens and observer are aligned, then the observer will see the light from the source smeared in a circle with the lens in its centre. Radius of the ring $\theta_E$ on the celestial sphere depends on the position and strength of the lens:
\begin{equation}
\theta_E = \sqrt{\frac{4GM}{c^2}\;\frac{d_{LS}}{d_L d_S}},
\end{equation}
where angular diameter distance $d_L$ is to the lens, $d_S$ is to the source, and $d_{LS}$ is between the lens and the source, see figure \ref{lens}. Meaning that, by measuring distance to the lens and to the source, one can estimate total mass, $M$, of the lensing object from the size of the Einstein ring. One should keep in mind that for angular distances $d_{LS}\ne d_S-d_L$ holds in general, while $d_L$ and $d_S$ are usually estimated from the measured redshifts and depend on cosmological model and its parameters. The first full ring ever observed is B1938+666, detected by HST in 1998 \cite{King98}. More recent results for the lensing galaxy of this system give $M_{lens}=2.46\times10^{10}M_\odot$, which is orders of magnitude larger than the visible mass in the ring. This is another valuable evidence for presence of DM in galaxies or clusters, as it has been confirmed with other observed rings, too. The Horseshoe system is another example of ring formation. In the centre of the ring sits just a single galaxy, but the radius of the ring is very large $R_E=30$kpc, suggesting an extremely massive DM halo \cite{Belokurov97}.\\
In more common case source, lens and observer are not exactly aligned, there is no ring formation, but the lens can produce multiple images of the source. However, even with alignment of the source and the lens, a large separation of source's images was observed in some rare cases. This would not have happened if the most of the lens' mass is concentrated in its centre. Instead, these rare cases are fitted well with the model that accounts for DM halo \cite{Maoz93}. Note that large statistics of strong lensing images was available years before the full Einstein ring was observed. With detection of multiple images there are additional observables (besides redshifts) that can be measured, for instance: the apparent coordinates of source's images and lens' coordinates, and also the time delay between the observed images. Measurements can be used to calculate true coordinates of the source, mass of the lens and distances to the lens and the sources. With large statistics strong lensing gives relation between angular diameter distance and the redshift, which then can be fitted to a cosmological model, and give estimate of cosmological parameters \cite{Nakamura96, Cao12}. Neighbourhood of lensing object may also contribute to the lensing. This, for example, can be the cluster in which lensing galaxy is located, or its satellite galaxies, which produce distortions in the lensed images. Analysis of distortions gives insight into matter distribution in and around the lens. Interestingly, there was an unconfirmed case of microlensing by a planet, occurring on the top of the strong lensing \cite{Schilling96}.
\subsection{Weak lensing}
Weak lensing is a phenomenon observed by statistical analysis of a number of background objects whose observed images are elongated due to gravitational lensing by large matter distribution located at the line of sight. Distortions of background are very small, hence large sets of data are required for statistical methods to be used, but on the other side, it is the most common observation of gravitational lensing. Weak lensing is used in various ways and it became one of the principal probes of dark matter, and even more general - of cosmological models.\\
Distortions of images of background galaxies are defined via relation between displacement 2D vectors in the image plane, $\delta x^I$ and in the source plane $\delta x^S$=A$\delta x^I$, where A is the distortion matrix:
\begin{equation}
A=\left(
\begin{tabular}{cc}
$1-\kappa-\gamma_1$ & $-\gamma_2$\\
$-\gamma_2$ & $1-\kappa-\gamma_1$
\end{tabular}
\right),
\end{equation}
where $\kappa$ stands for convergence, while $\gamma_1, \gamma_2$ define complex cosmic shear $\gamma=\gamma_1+i\gamma_2$. Cosmic shear actually measures tangential stretching of background sources as observed in plane perpendicular to the line of sight, along two chosen axes $\gamma_1, \gamma_2$. The method includes the measurement of the ellipticities of the background galaxies and construction of a statistical estimate of their systematic alignment. The fundamental idea counts on assumption that intrinsic ellipticities of observed galaxies average out to zero, but the presence of mass at the line of sight produces observable pattern of alignments in background galaxies, but with the distortion of only $\sim0.1\%-1\%$. Although previously described and predicted, the effect was long awaited and for the first time observed in 2000 \cite{Wittman00, Bacon00, Waerbeke00}. 
To show theoretical basics of weak lensing we write perturbed metric of the standard model, in Newtonian Gauge, as:
\begin{equation}
ds^2 =-(1+2\Psi)c^2dt^2+a^2(t)(1-2\Phi)\left[d\chi^2+r^2(d\theta^2+sin^2\theta d\varphi^2)\right],
\end{equation}
where we used $\chi$ for radial coordinate, while $r(\chi)$ is comoving distance, and $\Psi$ and $\Phi$ are gravitational potentials.
Cosmic shear and convergence are defined as:
\begin{align}
\begin{split}
\gamma&=\frac12(\psi_{,11}-\psi_{,22})+i\,\psi_{,12}\\
\kappa&=\frac12(\psi_{,11}+\psi_{,22})
\end{split}
\end{align}
where $\psi$ is the projected Newtonian potential, $\psi_{,ij}=-1/2\int d\chi\,g(\chi)\left(\Psi_{,ij}+\Phi_{,ij}\right)$, while commas are used to denote partial coordinate derivatives.
Secondly, $g(\chi)=r(\chi) \int_\chi^\infty d\chi' n(\chi')\frac{r(\chi'-\chi)}{r(\chi')}$, with $n(\chi)$ being the normalized radial distribution of source galaxies. Under assumption that all sources are at a single redshift it is
\begin{equation}
g(\chi)=r(\chi)r(\chi_s-\chi).
\end{equation}
Measuring the distortion matrix of set of objects in background of the lens can be used to map statistically the matter distribution in the foreground objects. This method has been used on individual lensing objects from 1990 \cite{Tyson90}. In this example Tyson et al. estimated matter distribution in two galaxy clusters by analyzing 20-60 faint background galaxy images, and found results in agreement with measurements of velocity dispersion and X-ray observations. 
More famous example, Bullet cluster 1E 0657-558, on the contrary, shows distinct offset of X-ray emitting hot gas from dark matter halos inferred by weak lensing method \cite{Markevitch04}. 
\begin{figure}
\center
\includegraphics[scale=0.7]{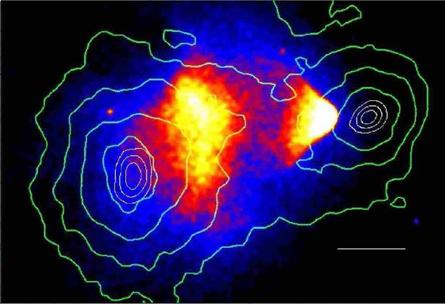}
\caption{Bullet cluster: X-ray image from Chandra (red); mass density contours (green) and DM (blue) are obtained through weak lensing \cite{Clowe06}.}
\label{bullet}
\end{figure}
This is a pair of galaxy clusters (figure \ref{bullet}), where the smaller ($7\times10^{13}{\rm M}_\odot$) subcluster (bullet) is just exiting the collision site, away from a $2\times10^{15}{\rm M}_\odot$ cluster, almost tangentially to the line of sight. A prominent bow shock gives an estimate of the subcluster velocity, $4500\pm1000$ km/s. The optical image shows that the gas lags behind the subcluster galaxies. The weak-lensing mass map reveals a dark matter clump lying ahead of the collisional gas bullet, but coincident with the effectively collisionless galaxies. The hot X-ray gas has been separated by ram pressure-stripping during the passage. This separation is only possible if the dominant mass is in the collisionless component, i.e. in the non-baryonic dark matter halo, not in the baryonic X-ray gas. From these observations, one can directly estimate the upper limit on cross-section of the dark matter self-interaction, which is of the order of $\sigma/m<0.7{\rm cm}^2$/g \cite{Randall08}.\\
Unlike cosmic shear, convergence can be easily related to cosmological model. Convergence in direction $\vec{n}$ is:
\begin{equation}
\kappa(\vec{n},\chi)=\int_0^\chi W(\chi')\delta(\chi')d\chi',
\label{conv-W}
\end{equation}
where the integral goes along the line of sight, $\delta$ is the relative perturbation in matter density and $W(\chi)$ is a weight function.
\begin{equation}
W(\chi)=\frac32\Omega_m H_0^2g(\chi)(1+z)
\label{W}
\end{equation}
To use this relation one needs large statistics of measured convergence in different directions in the sky. Individual convergences  cannot be theoretically estimated, but their correlations can. Two-point correlation function is the best exploited, but higher orders are analyzed as well \cite{Cooray01, Schneider05}. Power spectrum of convergence can be expanded in spherical harmonics:
\begin{equation}
\kappa_{lm}=\int d\vec{n}\,\kappa(\vec{n},\,\chi)Y_{lm}
\end{equation}
Assuming isotropic convergence function, its power spectrum is then defined from two-point correlation function:
\begin{equation}
\left<\kappa_{lm}\kappa_{l'm'}\right>=\delta_{l}\delta_{l'}\delta_{m}\delta_{m'}P^l
\end{equation}
In the case of weak distortions the power spectrums of convergence and cosmic shear are equal, and this one power spectrum is measured combining all available observable quantities. Radial dependance is usually included discontinuously by defining redshift bins (noted as $i,\,j$). This generalizes power spectrum:
\begin{equation}
\left<\kappa_{lm,\,i}\,\kappa_{l'm',\,j}\right>=\delta_{l}\delta_{l'}\delta_{m}\delta_{m'}P_{ij}^l
\end{equation}
From equation (\ref{conv-W}, \ref{W}) convergence power spectrum can be related to cosmology \cite{Kaiser98}:
\begin{equation}
P_{ij}^l=\int_0^\infty dz\frac{W_i(z)W_j(z)}{r(z)^2H(z)}P\left(\frac l{r(z)},z\right),
\end{equation}
where all functions are now adopted to using redshift bins in radial direction.\\
Finally, weak lensing statistics can be used to put constraints on cosmological parameters. Actually, the most lensing results constrain only the product $\sigma_8\Omega_m$, which brings a strong degeneration between the latter two parameters. \cite{Jarvis06, Semboloni06, Fu08}.
\section{Constraints on DM from cosmological observables}
Other valuable information on dark matter comes from data sets such as cosmic microwave background radiation (CMB), supernovae type Ia (SNIa) and baryonic acoustic oscillations (BAO), separately, and even better combined to give more stringent constraints. They measure dark matter density parameter on cosmological scales, bringing complementary information  to previously described methods that are primarily used on individual objects. Moreover, latter require background cosmological parameters (such as $\Omega_m,\,\Omega_\Lambda$ etc.) in order to evolve models of galaxy and cluster dynamics. In few words, local and cosmological observables form a strong feedback for our understanding of the Universe on every scale.
\subsection{Cosmic microwave background radiation}
Likely, the most powerful astrophysical observable for constraining DM is the cosmic microwave background radiation. These photons are in microwave regime and represent a fossil radiation coming directly from the last scattering surface (LSS), which is the epoch when cosmic matter and radiation decoupled. Its black body nature (Nobel prize in 1978 to Penzias and Wilson \cite{PenziasWilson65}) is one of the pillars of the standard cosmological model and was the definite conformation for the Big Bang theory. Actually, CMB has the best black body spectrum ever observed, with an average temperature $T_0=2.726$K \cite{Jarosik11}. If the measurement of "zero order" temperature of the CMB had been revolutionary for our understanding of the Universe, the higher orders that follow (and their discoveries)  would not have been less exciting. Anisotropy in CMB is expressed as a relative variation from average temperature $\Delta T/T$ measured across the sky, and it is directly related to total radiation energy by Stefan-Boltzmann law.\\
Small fluctuations in the cosmic matter distribution, which originate from the inflationary stage of the early Universe, were predicted to have been increased eventually by gravity, forming the large scale structures we observe today \cite{Peebles70}, \cite{Zeldovich70}. In the early, dense Universe, matter and radiation where strongly coupled via scattering interactions, meaning that perturbations in radiation distribution reflect perturbations in matter distribution as well. 
Relative perturbations in CMB of the order $10^{-4}-10^{-5}$ where discovered in 1992 by COBE  experiment (Nobel prize in 2006 \cite{Smoot92}), confirming the general picture of the large scale structures formation theory. COBE's discovery was not the end of the story. The era of precision cosmology was just started and it was reached little by little since the first trials of measuring angular power spectrum of CMB was measured.\\
Anisotropies come primarily from processes dating back to the era of recombination, but also from secondary processes that have happened in the eras that followed LSS, up to today. Detecting and studying both kinds of anisotropies provides us a lot of information about the early, but also later stages of cosmic evolution. Universe before LSS is modeled as multicomponent fluid with stronger interactions due to high density, and hence the components' perturbations were related. Connection is given by their adiabatic nature. In fact, adiabatic intrinsic perturbations in the early cosmic fluid naturally arise from the simplest inflationary scenario \cite{Riotto02}. In regard to perturbations we recall \cite{MaBertsinger95} the Jeans equation for the early cosmic fluid:
\begin{equation}
\ddot\delta_i+2\frac{\dot a}a\delta_i-\frac{c_s^2}{a^2}\Delta\delta_i=4\pi{\rm G} \sum_j\rho_j\delta_j,
\end{equation}
with $\delta_i$ being relative density perturbation of one of the components, like baryons or dark matter, $c_s$ is the sound speed, and $a$ is the scale factor; dots refer to time derivatives, while $\Delta$ is Laplacian operator.
The term on the right hand side sums up to total gravitational force of all matter components. Since DM contribution is by order of magnitude larger than other terms, they can be neglected. The equation can then be solved starting from DM component and then for others, too.
Fundamental effects that created CMB anisotropies at LSS are: $i$) Sachs-Wolfe effect, $ii$) intrinsic adiabatic perturbations, and $iii$) Doppler perturbations. Set of equations that describe the effects is in general more complicated \cite{MaBertsinger95, Challinor99}. They require to develop the perturbed Einstein equations coupled with Boltzmann equations. However, Jeans equation is derived from them and remained the back bone of the physics of primordial perturbations.
Sachs-Wolfe effect is based on the fact that matter density fluctuations on LSS, and hence gravitational field fluctuations affected photons escaping from the perturbed regions \cite{SachsWolfe67}. 
From one side, overdense region has stronger gravitational field that redshifts the escaping photons. From the other, the same region is overdense in photons as much, and appears more luminous, in opposition to previous. We thus expect the two terms to partially cancel \cite{White97}. The later one wins, but reduced with factor $\frac13$.\\
%Once radiation and matter decoupled, photons leaving out of overdensed regions on LSS were redshifted, while photons leaving out of underdensed regions gained energy and blueshift. However, this effect goes together with the second one. As said, fluctuations in matter density induce perturbations in photon density. That means the overdensed regions also contained proportionally more photons as much as baryonic matter. The higher density of photons should be observed as higher intensity of CMB radiation coming from that spot, giving it higher temperature. This is opposite to the effect of collective redshifting of the photons coming from the same region. We thus expect the two terms to partially cancel \cite{White97}. The later one wins, but reduced with factor $\frac13$.\\ When talking about primordial Sachs-Wolfe effect, the $\frac13$ factor is usually already considered. It turns out that Sachs-Wolfe effect is constant along different angular scales in CMB radiation, what makes it the predominant source of CMB fluctuations on large scales, above about ten degrees.
%Check why Sachs-Wolfe effect would be constant over scales, and for small scales why grav redshift would be related to photon overdensities since baryonic overdensity is much smaller than of the dark matter.
The Doppler effect was first explained by Sunyaev and Zel'dovich, and it does share the same mechanism with Sunyaev-Zel'dovich effect. Density perturbations in early cosmic fluid induce also velocity field in plasma. The last time when most of the CMB photons scattered off of matter was on LSS. Although average flow of this matter was zero, individual baryons did not have zero speed (in comoving frame of reference), inducing in this way temperature fluctuations in observed CMBR. Different from this, Sunyaev-Zel'dovich effect is created after LSS, by scattering of CMB photons along the line of sight, on hot plasma electrons, usually in intracluster medium. Thanks to this effect dense galaxy clusters have been observed.\\ %Besides Sunyaev-Zel'dovich effect, another secondary anisotropy effect in CMBR origins from integrated Sachs-Wolfe or Rees-Sciama effect. This effect is present because of difference in cosmic times of `falling in' and `going out of' gravitational potential wells along the line of sight. It is sensitive to the cosmological constant and arises just recently in cosmic history, in dark energy era.\\
To properly analyze CMB anisotropies, we have to look on what statistical level anisotropies of different scales occur, i.e. what is the power spectrum of CMB. For this task radiation field, or sky map, is decomposed into spherical harmonics $Y_{lm}$:
\begin{equation}
\frac{\Delta T}T(\theta,\varphi)=\sum_{l=0}^{\infty}\sum_{m=-l}^{m=l}a_{lm}Y_{lm}(\theta,\varphi)
\end{equation}
The main source of information is contained in angular power spectrum defined as 
\begin{equation}
C_l=\left<|a_{lm}|^2\right>=\sum_{m,m'=-l}^l a_{lm}^{}a_{lm'}^*.
\end{equation}
The assumption we use is that temperature fluctuations are Gaussian (which is not fully true), and $<..>$ represents an ensemble average over an infinite realization. Concretely, the theory predicts specific value of anisotropies that is randomly assigned to different directions. Since our Universe represents just one possible realization, we are unable to measure average $C_{lm}$. Instead we are only interested at average $C_l$ over $m$-modes, $C_l=\frac1{2l+1}\sum_m |a_{lm}|^2$. This measured value has also an unavoidable error $\Delta C_l=\sqrt{2/(2l+ 1)}$, called cosmic variance.\\
\begin{figure}
\center
\includegraphics[scale=0.5]{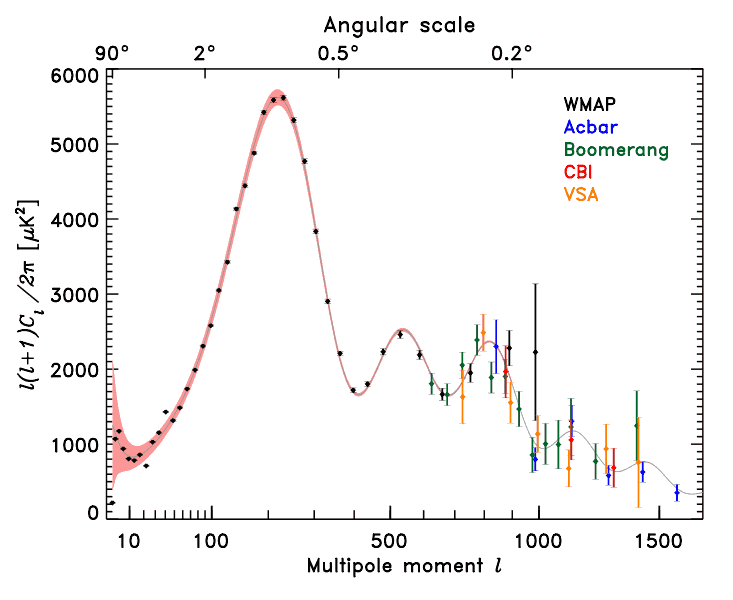}
\caption{CMB power spectra as estimated from several experiment. Solid line is the best fit model.}
\label{cmb}
\end{figure}
The locations of the acoustic peaks on figure \ref{cmb}. contain valuable information \cite{Efstathiou00}: $l_i=\alpha\,i\,\pi\,d_A^{\rm rec}/r_s$, where $\alpha$ is a number of order unity, $d_A^{\rm rec}$ is the angular diameter distance at the time of recombination:
\begin{align}
\begin{split}
d_A^{\rm rec}&=\frac{c}{H_0\left|\Omega_k\right|^{1/2}}\sin_k(\left|\Omega_k\right|^{1/2}x),\\
x&\approx\int_{a_r}^1\frac{da}{\left[\Omega_m a+\Omega_k a^2+\Omega_\Lambda a^4\right]^{1/2}}
\end{split}
\end{align}
For example, the angular scale of the first peak is strongly related to the curvature of the universe or $\Omega_k$ parameter. The odd numbered (first, third, fifth...) acoustic peaks are associated with how far the plasma ``falls" into gravitational potential wells (how much the plasma compresses), hence they are enhanced by an increase in the amount of baryons. On the contrary, the even peaks, show how far the plasma used to "rebound" (how much it rarefies). Therefore, the ration between odd and even gives a measure of the baryon density in the Universe, or $\Omega_b$ parameter. The third peak is sensitive to the total matter density $\Omega_m$ and so on.\\
Certainly, when computing final values of cosmological parameters the whole power spectrum, or the full data set, is used in much more complicated procedure that requires maximum likelihood sophisticated packages that sample smartly the parameter space (e.g. Gibbs sampling).
In the standard spatially-flat six-parameter $\Lambda$CDM cosmology Planck data determine the cosmological parameters to high precision \cite{Planck13}. In one sigma region they are: $\Omega_m=0.315\pm0.017$, $H_0=(67.3\pm1.2){\rm km\,s^{-1}Mpc^{-1}}$, $\Omega_b h^2=0.02205\pm0.00028$, $\Omega_c h^2=0.1199\pm0.0027$ etc.
\subsection{Combining CMB with other data sets}
Better constraints to the DM content in the Universe can be obtained combining different techniques, like BAO and SNIa with CMB \cite{Amanullah10}. Theory of large scale structure formation requires anisotropies in the CMB, but also the presence of dark matter in early Universe. Direct observations of large scale structures and its distributions is also used to put constraints on dark matter density \cite{Springel05}.
\subsubsection{Supernovae Ia}
\begin{figure}
\center
\includegraphics[scale=0.7]{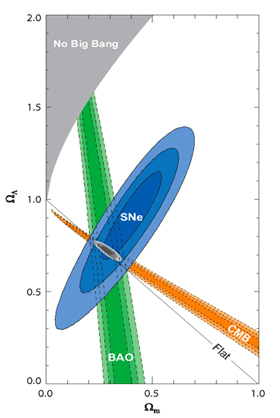}
\caption{$1\sigma,\,2\sigma$ and $3\sigma$ confidence regions on plane $\Omega_m-\Omega_\Lambda$ from CMB, SNIa and BAO data \cite{Amanullah10}}
\label{ell}
\end{figure}

Supernovae (SN) are extremely luminous explosions of dying stars. This makes them directly observable even at very far distances. Classification of SN was originally done spectroscopically, but even this simple identification tells a lot of information about the star's evolution and final explosion. The most important type and the one relevant for cosmological measurements are supernovae Ia. These explosions happen from collapsing white dwarfs in close binary star systems. The explosion is triggered when white dwarf reaches the Chandrasekhar mass limit in process of accretion from the binary companion \cite{Chandra83}. Since the limit has little varying value, all SN Ia are considered to have quasi-equal peak absolute magnitude. They are successfully identified both by spectrum and by the light curve. Luckily SN Ia is the most luminous and the most frequent type of supernovae explosions in the Universe.\\
By measuring their apparent magnitude, and joint with the redshift, this becomes a powerful method for sampling luminosity distance versus redshift relation. That function depends on cosmological model and can be used to constrain cosmological density parameters. Under standard model, the measurements reveal accelerated expansion of the Universe that is explained by contribution of dark energy \cite{Riess98, Perlmutter99}. Although primary result is existence of dark energy, measurement also bounds the value of total matter density. 
The two parameter space is degenerate over deceleration parameter $q =-\ddot{a} a /\dot{a}^2$ that is fortunately in an orthogonal direction with respect to confidence regions coming from CMB data (see figure \ref{ell}).
\subsubsection{Baryonic Acoustic Oscillations}
As introduced above, early cosmic fluid was driven by the gravitational force of matter and pressure of photons. The competition  between the two provoked oscillations in form of the sound waves. These are called baryonic acoustic oscillations (BAO). At the time of recombination photons decoupled, while baryons, not supported by the pressure, started forming dense structures by gravitational bonding. Sound horizon at the time of LSS is measurable from the CMB data. After that, matter interacted gravitationally and Universe expanded. However, current distribution of large structures provides measure of the sound horizon today. Just like supernovae Ia are standard candles of the Universe, BAO sound horizon is the standard ruler. Expansion of the Universe is tracked by evolution of this standard ruler and can be compared to theoretical models.
New combined results from Supernova Cosmology Project, have the best fit values related to the figure \ref{ell}: $\Omega_m=0.281_{-0.016}^{+0.018},\,\Omega_k=-0.005\pm0.007$ \cite{Amanullah10}, which seems to correspond exactly to the flat model.
%\subsubsection{Nucleosynthesis}
%Direct determinations as well as the nucleosynthesis constraints show that the density of baryonic matter is only about 4\% of the critical density ($\Omega_b=0.04$). There must exist some other forms of matter than ordinary matter.
%\section{Conclusion}
%Fermi Gamma-ray Space Telescope, launched on June 11, 2008, detected gamma rays with peaks at energies at 111 and 129 GeV in the Galactic center. The double peak-like excess can be interpreted as a signal of dark matter direct two-body annihilations into two channels with monochromatic final-state photons \cite{Weniger12}. Similar result comes from other clusters \cite{Tempel12a},  \cite{Hektor13}.
\section{Conclusions}
It has been almost one century since the dark matter paradigm has been introduced. Nowadays various fields in physics confirm the dark matter existence, and astrophysics is very fervent in respect to that, as we have seen. We have discussed proofs for dark matter from the scale of a galaxy to the cosmological one, such us galaxies' rotational curves and CMB power spectrum, respectfully. We would like to emphasise that although the error bars on the value of dark matter parameter have been shrinking significantly, latest results from Planck experiment show a bit of discrepancy with respect to the earlier accepted value (see the last section). This means today's focus is not on the existence of dark matter, but on the precise measurements of its properties and nature. With respect to this a lot of fresh information is expected from the next generation experiments in galaxy surveys, such as Euclid \cite{Euclid12} and particle physics. The former one is a mission specifically dedicated for measuring galaxy shapes and their corresponding redshifts, taking advantage of a powerful weak lensing techniques, and aims to get strong constraints on dark matter comparable to the current achievements.

\bibliography{Dark_matter_Lukovic_Cabella_Vittorio}{}
\bibliographystyle{ws-ijmpa}

\end{document}